\documentclass[aps, pra, superscriptaddress, a4paper, showpacs, twocolumn, english, 10pt]{revtex4-2}
\usepackage{bbm, amsthm, bm,textcomp, nicefrac,geometry,ragged2e}
\usepackage[dvipsnames]{xcolor}
\usepackage{float}
\usepackage[bbgreekl]{mathbbol}
\usepackage{graphicx,epstopdf,color,verbatim,enumitem}
\usepackage{pbox,array}
\usepackage{mathrsfs}
\usepackage{amssymb}
\usepackage{textgreek}
\usepackage{mathtools}
\usepackage{stackrel}
\usepackage[thinlines]{easytable}
\usepackage{amsmath}
\usepackage[utf8]{inputenc}

\makeatletter
\usepackage{soul}
\usepackage[normalem]{ulem}
\usepackage[caption=false]{subfig}
\usepackage{hyperref}
\usepackage[T1]{fontenc}
\usepackage[caption=false]{subfig}
\usepackage{babel}
\usepackage[linesnumbered,ruled]{algorithm2e}

\hypersetup{
    colorlinks = true,
    linkcolor = purple,
    citecolor = purple,
    filecolor = purple,      
    urlcolor = purple,
}

\newcommand{\prot}{SQEM}

\newcommand{\ket}[1]{\left|#1\right\rangle}

\newcommand{\bra}[1]{\left\langle #1\right|}

\newcommand{\proj}[1]{\ket{#1}\!\bra{#1}}

\newcommand{\id}{\mathbb{1}}

\newcommand{\tr}{\text{Tr}}

\begin{document}

\title{Superposed Quantum Error Mitigation}

\author{Jorge Miguel-Ramiro}
    \thanks{These authors contributed equally.}
    \affiliation{Universit\"at Innsbruck, Institut f\"ur Theoretische Physik, Technikerstra{\ss}e 21a, 6020 Innsbruck, Austria}
\author{Zheng Shi}
    \thanks{These authors contributed equally.}
    \affiliation{Institute for Quantum Computing, University of Waterloo, Waterloo, ON N2L 3G1, Canada}
    \affiliation{Department of Physics \& Astronomy, University of Waterloo, Waterloo, ON N2L 3G1, Canada}
\author{Luca Dellantonio}
    \affiliation{Institute for Quantum Computing, University of Waterloo, Waterloo, ON N2L 3G1, Canada}
    \affiliation{Department of Physics \& Astronomy, University of Waterloo, Waterloo, ON N2L 3G1, Canada}
    \affiliation{Department of Physics and Astronomy, University of Exeter, Stocker Road, Exeter EX4 4QL, United Kingdom}
\author{Albie Chan}
    \affiliation{Institute for Quantum Computing, University of Waterloo, Waterloo, ON N2L 3G1, Canada}
    \affiliation{Department of Physics \& Astronomy, University of Waterloo, Waterloo, ON N2L 3G1, Canada}
\author{Christine A. Muschik}
    \affiliation{Institute for Quantum Computing, University of Waterloo, Waterloo, ON N2L 3G1, Canada}
    \affiliation{Department of Physics \& Astronomy, University of Waterloo, Waterloo, ON N2L 3G1, Canada}
    \affiliation{Perimeter Institute for Theoretical Physics, Waterloo, Ontario N2L 2Y5, Canada}
\author{Wolfgang D\"ur}
    \affiliation{Universit\"at Innsbruck, Institut f\"ur Theoretische Physik, Technikerstra{\ss}e 21a, 6020 Innsbruck, Austria}

\begin{abstract}
Overcoming the influence of noise and imperfections is a major challenge in quantum computing. Here, we present an approach based on applying a desired unitary computation in superposition between the system of interest and some auxiliary states. We demonstrate, numerically and on the IBM Quantum Platform, that parallel applications of the same operation lead to significant noise mitigation when arbitrary noise processes are considered. We first design probabilistic implementations of our scheme that are plug and play, independent of the noise characteristic and require no postprocessing. We then enhance the success probability (up to deterministic) using adaptive corrections. 
We provide an analysis of our protocol performance and demonstrate that unit fidelity can be achieved asymptotically. Our approaches are suitable to both standard gate-based and measurement-based computational models.
\end{abstract}
\date{\today}
\maketitle

\textit{Introduction.---} Quantum computers can solve problems that are not accessible by classical devices \cite{national2019quantum,Baker2021}, ranging from factoring large numbers to applications in quantum chemistry \cite{NAP25196,Paulson2021Sim,Haase2021res}. However, noise and imperfections restrict practical applications \cite{Preskill2018,Bharti2022}. Advanced, resource intensive methods such as quantum error correction \cite{GottesmanThesis,Lidiar2013} or fault-tolerance \cite{ShorFault,GottesmanClifford} are expected to overcome these limitations. Yet, stringent error thresholds must be met alongside large numbers of required qubits, thereby making such approaches challenging in the short-medium term.

Here, we propose an alternative method to reduce noise in quantum gates and circuits. Our approach, called Superposed Quantum Error Mitigation (SQEM), is based on applying quantum gates in a superposed, coherently controlled way, on either the input state or some auxiliary system. A measurement of a control register and the auxiliary system leads to a probabilistic enhancement of gate fidelities. At the cost of an additional calibration, the success probability of {\prot} can be enhanced up to becoming deterministic. 

The method we introduce is similar in spirit to the superposition of paths \cite{Chiribella2019,Kristjnsson_2020,Abbott2020,MR2021} and that of causal orders \cite{Procopio2015,Ebler2018,Guerin2019,Caleffi2020,Guo2020,Chiribella2021}, which are advantageous in computation \cite{Araujo2014} and communication \cite{Gisin_2005, Araujo2014, Procopio2015, Abbott2020, Chiribella2019, Caleffi2020, Chiribella2021, Rubino2021}. However, {\prot} employs controlled-SWAP (cSWAP) operations, similar to \cite{MiguelRamiro2021,Huggins2021,Koczor2021}. It is conceptually easier to understand, implement and analyze, providing stronger advantage and addresses the main drawbacks of the other approaches.  In fact, SQEM has proven advantageous on an IBM Quantum device, confirming that it works with noisy control registers and cSWAP operations. The desired gate or circuit simply needs to be independently applied to several subsystems, after producing the required superposition. Noise operators destructively interfere, thereby enhancing the output fidelity. Surprisingly, this does not only happen probabilistically for a few measurement outcomes. With appropriate correction operations, deterministic advantage is obtained. Finally, SQEM is not limited to the correction of estimated observables as in \cite{Temme2017,Huggins2021,Koczor2021}. Instead, it yields a quantum state that can be further processed in subsequent computations, and can be applied for different purposes such as enhancing quantum memories (see also \cite{pra_us}). Moreover, SQEM only requires a single copy of the input state and is resilient against noise affecting the additional operations required for its operation.

In the simplest case, {\prot} involves two cSWAP operations and two applications of the desired gate. While our protocols work with any gate, here we focus on the cNOT and the non-Clifford T gates. The approach can be scaled up, either for whole computations in superposition on many input qubits or applying individual gates multiple times on large auxiliary systems. In the latter situation, it is possible to asymptotically obtain noiseless gate implementations. Remarkably, the underlying computational model is largely irrelevant. We demonstrate that for both gate-based (GB) \cite{nielsen_chuang_2010} and measurement-based (MB) \cite{Briegel2001,Oneway2005,Briegel2009} quantum computation (QC) fidelities are enhanced.

While in the GB approach adding control leads to a direct superposition of noise processes, in the MB model static noise from imperfect preparation of resource states is superposed by means of cSWAP performed before and after the application of gates. For MB-QC all operations, including the cSWAP, are realized by performing sequences of (possibly adaptive) single-qubit measurements on an entangled resource state. The cSWAP can be realized by different means, including via additional degrees of freedom naturally available in the physical information carrier \cite{Friis14}, and may themselves be noisy. Even if the noise levels of cSWAP and other gates are similar, one still finds an advantage in using {\prot}.

\begin{figure}
    \centering
    \includegraphics[width=\columnwidth]{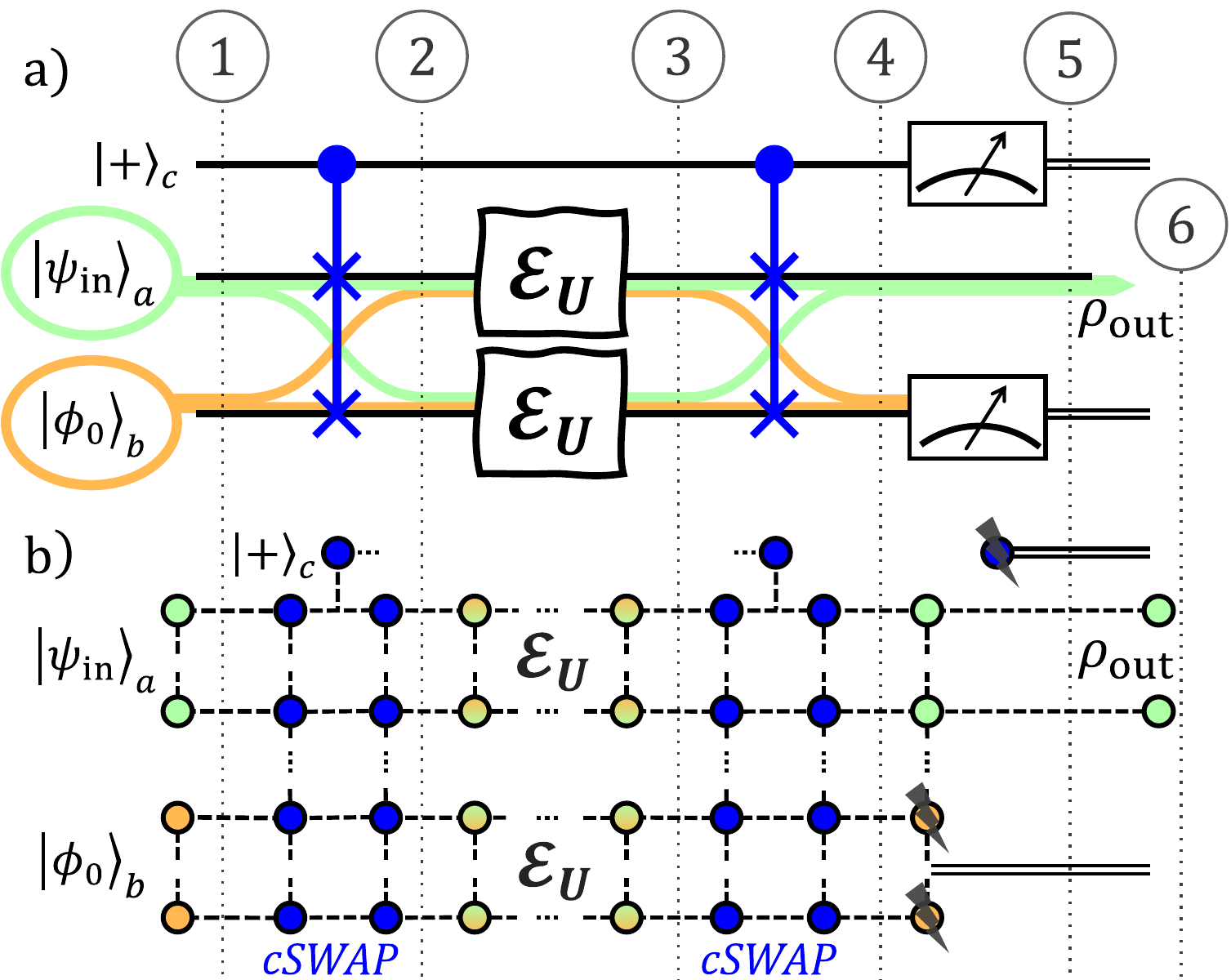}
    \caption{ Illustration of the strategy for enhancing the fidelity of noisy gate-based (a) and measurement-based (b) computations. The protocol steps from one to six (described in the main text) are highlighted. The input (auxiliary) state follows the superposed green (orange) paths depending on the state of the control. 
    }
    \label{fig:compstandard1_prl}
\end{figure}

\textit{Setting.---}
As schematically represented in Fig.~\ref{fig:compstandard1_prl}(a), we consider an $m$-qubit register ``a'' initialized in the input state $\ket{\psi_{\rm in}}_{\rm a}$, and an arbitrary computation $U$ that is subjected to noise. Noise is modelled \footnote{We remark that our protocols are not restricted to this particular noise model, which is considered for clarity. Our analytical results are independent of the noise model provided the Kraus operators are properly redefined.} by a channel acting after a perfect application of $U$, described by
\begin{equation}
\label{eq:gen_noise_gate_prl}
    {\cal E}_{U} (\rho) = \sum_j K_j \big( U \rho U^{\dagger} \big) K_j^{\dagger},
\end{equation}
where $\rho$ is a density matrix (e.g., $\proj{\psi_{\rm in}}_{\rm a}$) and $\{K_j\}$ are the Kraus operators associated with the noise. Without loss of generality, we set $K_0 = \sqrt{p_{\rm ne}} \id$ with $p_{\rm ne}$ being the probability of having no errors. 

The goal is to devise a protocol that can (partially) correct the noise affecting the operation $U$. To do so, we include two additional registers called control and auxiliary (same size as input), indicated with subscripts ``c'' and ``b'', respectively. As shown in Fig.~\ref{fig:compstandard1_prl}(a), the desired computation is implemented in superposition, such that $U$ acts simultaneously on \textit{both} the input $\ket{\psi_{\rm in}}_{\rm a}$ \textit{and} the auxiliary $\ket{\phi_{0}}_{\rm b}$ states. To achieve this, we swap the ``a'' and ``b'' registers depending on the state of the control. For $\ket{1}_{\rm c}$ ($\ket{0}_{\rm c}$), $\ket{\psi_{\rm in}}_{\rm a}$ and $\ket{\phi_{0}}_{\rm b}$ exchange (follow their own) branches. A ``branch'' is associated with each state ($\ket{0}_{\rm c}$ or $\ket{1}_{\rm c}$) of the control register, and corresponds to the path followed by the corresponding input state without the swapping.

Implementing the computation in a coherent superposition creates interference of the noise associated with $U$. Since $U$ acts on both the input and auxiliary states in \textit{each} branch, the noise becomes entangled. Later measurements of the control and auxiliary registers collapse the state such that specific errors are suppressed or enhanced, depending on the measurement outcomes. Based on these, one can then post-select the result or perform unitary corrections. In both cases, the fidelity of the output state $\rho_{\rm out}$ is enhanced compared to the incoherent case ${\cal E}_{U}(\proj{\psi_{\rm in}}_{\rm a})$ [see Eq.~\eqref{eq:gen_noise_gate_prl}].

To quantify the advantage of {\prot}, we employ the Choi-Jamiołkowski (CJ) fidelity $F_{\rm CJ}$ \cite{jozsa1994fidelity, Glischrist2005} with respect to a perfect implementation of $U$,
\begin{equation}
\label{eq:CJfid}
  F_{\rm CJ}= \bra{\Phi_{m}^{+}} (\id \otimes U^{\dagger}) \rho^{\rm CJ}_{\rm out}  (\id \otimes U)\ket{\Phi_{m}^{+}}.
\end{equation}
Here, $\ket{\Phi_{m}^{+}}=(\ket{00}+\ket{11})^{\otimes m}/\sqrt{2^m}$ describes $m$ maximally entangled pairs of qubits (i.e., Bell states \cite{nielsen_chuang_2010}). We keep half of these --- the first qubit in each pair --- and use the remaining half as the input to perform $U$, yielding the output $\rho^{\rm CJ}_{\rm out}$.

$F_{\rm CJ}$ in Eq.~\eqref{eq:CJfid} is a lower bound on the achievable fidelity with a generic input state $\ket{\psi_{\rm in}}_{\rm a}$. With a formal demonstration in  \cite{Dur2005}, the idea is that \textit{any} noise acting on the chosen half of $\ket{\Phi_{m}^{+}}$ is maximally detrimental, as it destroys the entanglement of the composite density matrix and hence its coherence. Thus, we employ $F_{\rm CJ}$ for characterizing {\prot}. 

\textit{Protocol.---}
Below, we introduce and explain {\prot} for error mitigation in both GB- and MB-QC. For clarity, we consider two branches, i.e., only one auxiliary state $\ket{\phi_0}_{\rm b}$. The generalization to more branches and higher-level systems is in  \cite{pra_us}, where we also investigate variations of our schemes, provide extended numerical results and introduce an interferometric-based approach where the branches are physically represented by the arms of a multi-input/output interferometer. 

As schematically represented in Fig.~\ref{fig:compstandard1_prl}, our protocols to mitigate noise comprise the following steps:
\begin{enumerate}[leftmargin=\parindent,align=left,labelwidth=\parindent,labelsep=2pt]
\itemsep 0.2em

\item Prepare the control and auxiliary registers in $\ket{+}_{\rm c}= \left(\ket{0}_{\rm c}+\ket{1}_{\rm c}\right) / \sqrt{2}$ and $\ket{\phi_0}_{\rm b}$, respectively. 
\label{prot:step1}

\item Apply a cSWAP operation \cite{Fredkin82,pra_us} 
$\text{cSWAP} = \proj{0}_{\rm c} \otimes  \id  + \proj{1}_{\rm c} \otimes \text{SWAP}_{{\rm a}, {\rm b}}$
to coherently exchange registers ``a'' and ``b'' depending on the control register. 
\label{prot:step2}

\item 

Implement ${\cal E}_{U}$ in Eq.~\eqref{eq:gen_noise_gate_prl} in both registers ``a'' and  ``b'' independently.
\label{prot:step3}

\item Apply a second cSWAP operation as in step~\ref{prot:step2}.
\label{prot:step4}

\item Measure the control and auxiliary registers in the Pauli $X$ basis and an appropriate basis, respectively. The latter is chosen, based on $\ket{\phi_0}_{\rm b}$ and $U$, to maximize the fidelity of the output state $\rho_{\rm out}$ (see main text).
\label{prot:step5}

\item Depending on the measurement outcomes in step~\ref{prot:step5}, either post-select (``probabilistic'' variant) or post-select and correct (``quasi-deterministic'' variant) the output $\rho_{\rm out}$ in register ``a''.
\label{prot:step6}
\end{enumerate}

Below, we first describe the working principle behind our scheme. Afterwards, we characterize the probabilistic and the quasi-deterministic implementations in step~\ref{prot:step6}, investigating their advantages in realistic experimental settings. Finally, we analytically prove that our probabilistic protocol is always advantageous compared to the incoherent case 
${\cal E}_{U}(\proj{\psi_{\rm in}})$.

The state $\rho_{\rm out}$ after step~\ref{prot:step5} of {\prot} is (see also \cite{pra_us})
\begin{widetext}
\begin{equation}\label{eq:gate_out_prl}
    \rho_{\rm out} = \frac{\mathcal{A}}{2}
    \left[
        {\cal E}_{U} (\proj{\psi_{\rm in}}_{\rm a})
        \pm 
        \sum_{i,j} 
        \left(
        \frac{
        \bra{\phi_{\rm f}} K_j U \proj{\phi_{ 0}} U^{\dagger} K_i^{\dagger} \ket{\phi_{\rm f}}
        }{
        \mathcal{A}
        }
        \right)
        K_i U \proj{\psi_{\rm in}}_{\rm a} U^{\dagger} K_j^{\dagger}
        \right],
\end{equation}
\end{widetext}
where $\mathcal{A} = \sum_{i} \left\lvert \bra{\phi_{\rm f}} K_i U \ket{\phi_{ 0}}\right\rvert^2$ is a normalization constant and $\ket{\phi_{\rm f}}$ is the state onto which the auxiliary subsystem is projected in step~\ref{prot:step5}. The sign $\pm$ in Eq.~\eqref{eq:gate_out_prl} depends on the measurement outcome of the control register, with $+$ ($-$) corresponding to $\ket{+}_{\rm c}$ ($\ket{-}_{\rm c}$). The trace $\tr \left( \rho_{\rm out} \right)$ is the probability to obtain the auxiliary and control subsystems in the corresponding states. 

From Eq.~\eqref{eq:gate_out_prl} it is possible to understand how {\prot} works. The first term on the right-hand side describes the input state $\ket{\psi_{\rm in}}_{\rm a}$ always remaining in branch $\ket{0}_{\rm c}$, and thus resembles the incoherent case in Eq.~\eqref{eq:gen_noise_gate_prl}. Noise interference is found in the second, more interesting term.
The factor in the parentheses indicates that the larger the overlap of $\ket{\phi_{\rm f}}$ and $U \ket{\phi_{ 0}}$ is, the more $U\ket{\phi_{ 0}}$ is affected by the noise, and therefore the better our protocol works. 

This rather counter-intuitive fact is understood thinking in terms of the noise. After step~\ref{prot:step2}, the input and the auxiliary states are in a coherent superposition, and as such (in each branch) they are subjected to the same noise. The errors become correlated, and by measuring the control and auxiliary subsystems in step~\ref{prot:step5} we can learn about the noise that acted on
$U\ket{\psi_{\rm in}}$. The available knowledge increases when $U\ket{\phi_{0}}$ is more affected by the associated Kraus operators $K_i$ for $i\geq 1$ (i.e., it is orthogonal to their eigenvectors), and can be accessed if
$\ket{\phi_{\rm f}}$ is parallel to $U \ket{\phi_{ 0}}$. 

From these observations, to quantify the noise mitigation obtained with {\prot} we introduce 
\begin{equation}\label{eq:omegas_prl}
    \omega_1  
    = 
    1 -  \frac{ \sum_{j \geq 1} \left\lvert\bra{\phi_{ 0}} U^{\dagger} K_j U \ket{\phi_{ 0}} \right\rvert^2}{1-p_{\rm ne}} 
    , \,\,\,
    \omega_2 
    =
    |\bra{\phi_{\rm f}} U \ket{\phi_0} |^2
    ,
\end{equation}
where $\omega_1,\omega_2 \in [0,1]$. For $\omega_1 = 0$ ($\omega_1 = 1$) we say that $U \ket{\phi_{ 0}}$ is completely insensitive (sensitive) to all Kraus operators, and the correlations between noises affecting input and auxiliary states are minimized (maximized). 
Therefore, the extreme points $(\omega_1, \omega_2)=(1,1)$ and  $(\omega_1, \omega_2)=(0,0)$ correspond to maximal or minimal mitigation of the error affecting $U$, respectively. Any other pair of values of $(\omega_1, \omega_2)$ indicates a certain degree of noise mitigation and the corresponding advantage of {\prot}.

As mentioned in step~\ref{prot:step6}, the output $\rho_{\rm out}$ depends on the chosen implementation: the probabilistic or the quasi-deterministic. The first involves post-selection depending on the measurement outcomes at step~\ref{prot:step5}. This desired result includes the projection of the control register onto $\ket{+}_{\rm c}$, but depends on the chosen measurement basis for the auxiliary subsystem. Specifically, the state $\ket{\phi_{\rm f}}_{\rm b}$ must maximize $\omega_2$ in Eqs.~\eqref{eq:omegas_prl}. Ideally, $\ket{\phi_{\rm f}} = U\ket{\phi_0}$ such that $\omega_2 = 1$. In several experimental scenarios (e.g., $U$ being a Clifford circuit or $\ket{\phi_{0}}$ an eigenstate of $U$) this can be practically achieved. For simplicity, we consider this scenario in the following, albeit lower values of $\omega_2$ are not detrimental to the success of our schemes \cite{pra_us}.

At the cost of performing a calibration routine, i.e., repeated experiments to determine the correcting unitaries, 
the quasi-deterministic protocol enhances the post-selection probability of keeping $\rho_{\rm out}$. If required, this scheme works deterministically, i.e., without requiring post-selection. The idea is to employ a black-box optimization to find the best correcting unitaries to be applied to $\rho_{\rm out}$ depending on the measurement outcomes at step~\ref{prot:step5} \cite{pra_us}. This is done by repeated experiments where the different outputs $\rho_{\rm out}$ are analyzed and post-processed. While the probabilistic approach could be particularly useful for increasing the fidelity of whole computations in a plug-and-play fashion, the quasi-deterministic variant could be advantageous for optimizing one or a few gates that are repeated within a larger circuit, where the user can specify the desired success probability.

Below, we analytically quantify the advantage of our protocol. To better appreciate the potential of {\prot}, we present the following results when $(d-1)$ auxiliary branches are employed. The underlying idea is the same, except that now the input state $\ket{\psi_{\rm in}}_{\rm a}$ is (conditionally) swapped with $(d-1)$ identical auxiliary states $\ket{\phi_0}$ instead of one. Quantitatively, this means that the second term on the rhs of Eq.~\eqref{eq:gate_out_prl} is enhanced by the number $d$ of possible paths the input takes. The computational resources required for generalizing our protocols to $d$ branches are $(d-1) m$ qubits for the auxiliary registers and $\log_2(d)$ for the control. The discussion above (incoherent case) then refers to the case $d=2$ ($d=1$).

Despite the challenge of analyzing {\prot} in general settings, it is possible to derive theoretical results that are applicable to different experimental scenarios. Here, we consider the probabilistic implementation and $\omega_1 = 1$ (see \cite{pra_us} for a general analysis). A sufficient condition for $\omega_1 = 1$ is that $\bra{\phi_{\rm f}} K_j U \ket{\phi_0} = \sqrt{p_{\rm ne}} \delta_{j,0}$ for all $j$, where $\delta_{j,0}$ is the Kronecker delta. As explained above, this means that $U\ket{\phi_0}$ is maximally sensitive to the noise (recall $K_0 = \sqrt{p_{\rm ne}} \id$). In practice, this can be always achieved by employing a Choi-like state as auxiliary, i.e. using half of a Bell state and measuring in the Bell basis afterwards, see \cite{pra_us}.

Under these assumptions Eq.~\eqref{eq:gate_out_prl} becomes $\rho_{\rm out} = {\cal E}_{U} (\proj{\psi_{\rm in}}) + (d-1) p_{\rm ne} U \proj{\psi_{\rm in}}U^{\dagger}$, up to the normalization factor $1 + (d-1) p_{\rm ne}$. The associated CJ fidelity in Eq.~\eqref{eq:CJfid} is $F_{\rm CJ} = d p_{\rm ne}/[1+(d-1) p_{\rm ne}]$, which is a lower bound for the fidelity associated with an arbitrary input $\ket{\psi_{\rm in}}_{\rm a}$. Comparing $F_{\rm CJ}$ with the incoherent one $p_{\rm ne}$, we draw two important conclusions. First, $F_{\rm CJ}$ is a monotonically increasing function of $d$. This means that {\prot} always yields higher fidelity compared to the incoherent case, and by increasing $d$ we further enhance the result. Second, in the asymptotic case $d \gg 1$ we obtain a perfect implementation of $U$, regardless of the noise. 

The situation is more complicated when the chosen auxiliary state $\ket{\phi_0}$ is not maximally sensitive to the noise, i.e., $\omega_1 < 1$. In this scenario, it is still possible to demonstrate that the probabilistic protocol is always advantageous. Furthermore, the CJ fidelity increases with $d$ for sufficiently large values of $p_{\rm ne}$. However, for $d \gg 1$, $F_{\rm CJ}$ is limited to a value that depends on $\omega_1$ and that is lower than one. 
It is then possible to employ different auxiliary states to design iterative variations of our schemes to further enhance the output fidelity, see  \cite{pra_us}.

\textit{Results.---}
\begin{figure}[t]
    \centering
    \includegraphics[width=\columnwidth]{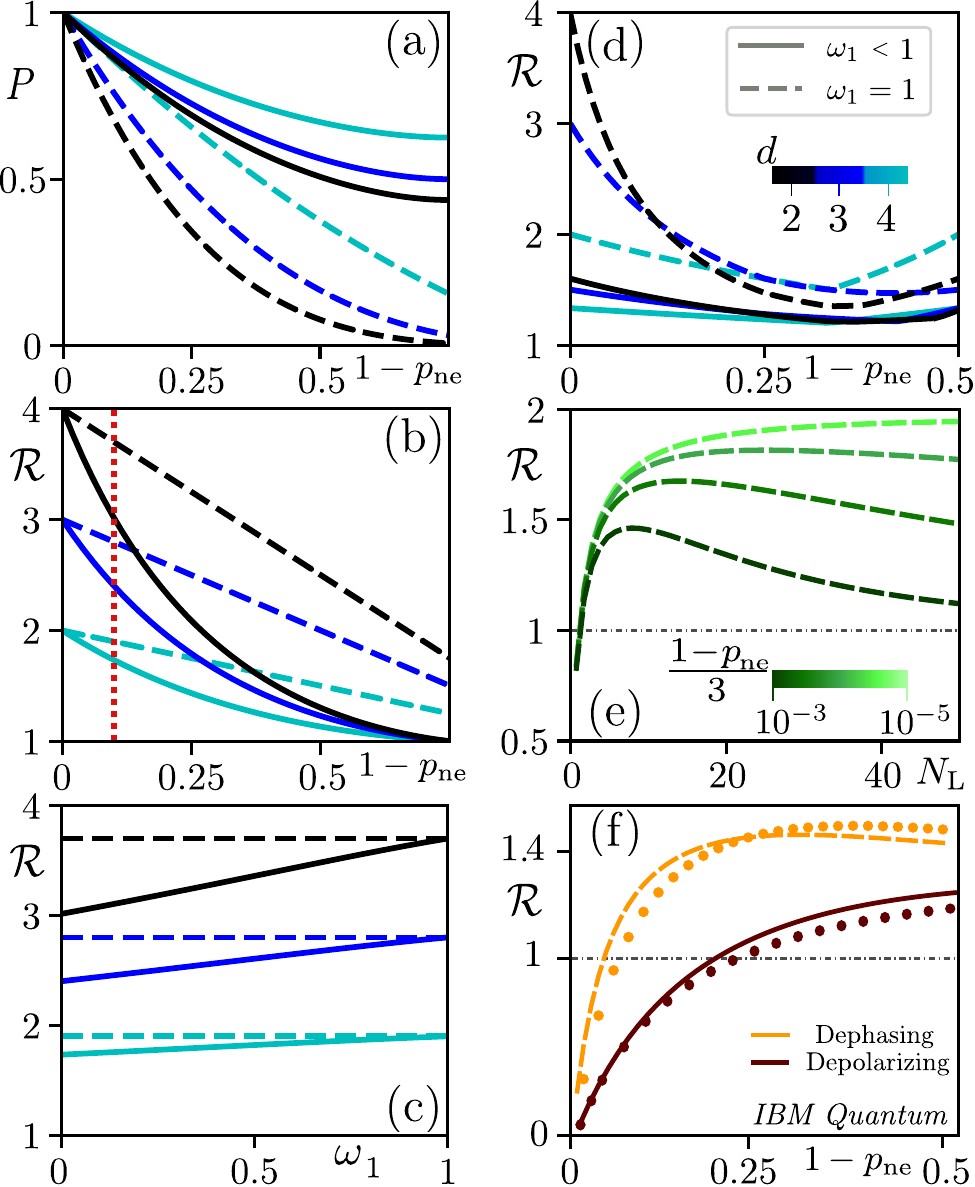}
    \caption{Post-selection probability $P$ and infidelity ratio $\mathcal{R}$ in different scenarios. Full and dashed lines are characterized by $\omega_1 < 1$ and $\omega_1 = 1$, respectively. In (a-c), the probabilistic scheme is considered, $U=\text{cNOT}$ followed by dephasing noise.
    In panel (c) we set $p_{\rm ne} = 0.9$ [dotted, red line in (b)] and show $\mathcal{R}$ for varying $\omega_1$.
    In (d), the deterministic protocol is applied to a T gate affected by dephasing. 
    In (e), the detrimental contribution of noise from the cSWAP is investigated for $\omega_1 = 1$. We consider the MB implementation of the probabilistic protocol for $U=[\text{cNOT}(T\otimes T)]^{N_{\rm L}}$, and
    depolarizing noise with error probabilities $1-p_{\rm ne}$ affecting each qubit of the resource state (including the ones implementing the cSWAPs). In (f), SQEM is demonstrated on the quantum computer ``ibm\_perth'', with $\ket{ \psi_{\rm in}}=\ket{ \phi_{0}}=\ket{+}$ and $U=\id$. Dots correspond to real device data (error bars are negligible compared to the size of the dots), while the lines represent the expected simulated behavior obtained by modeling each cNOT within the two cSWAP gates as a perfect operation followed by a depolarizing channel with noise parameters extracted from experimental data (see main text).}
    \label{fig:results_prl}
\end{figure}
In Fig.~\ref{fig:results_prl} we provide numerical and experimental results that confirm the analytical derivations above and demonstrate the advantage of our probabilistic and quasi-deterministic protocols in different settings. We set $\omega_2 = 1$ and identify $P$ and $\mathcal{R} = (1-F^{0}_{\rm CJ})/(1-F_{\rm CJ})$ as figures of merit, where $F^{0}_{\rm CJ}=p_{\rm ne}$ is the incoherent CJ fidelity. The parameter $P$ is the post-selection probability associated with the desired outcome(s) at step~\ref{prot:step5}. $\mathcal{R}$ quantifies the advantage of our schemes, such that for $\mathcal{R} \geq 1$ {\prot} is beneficial.

In panels (a-c) we consider the probabilistic scheme applied to a $U = \text{cNOT}$ gate. A dephasing channel acts independently upon each qubit, and the auxiliary state is set such that it varies $\omega_1$ in Eq.~\eqref{eq:omegas_prl} between zero and one. In the latter case, the analytical results above hold and we find $\mathcal{R} = 1+(d-1) p_{\rm ne}$ and $P =  p_{\rm ne}^{ d} \left[ 1 + \left( p_{\rm ne}^{-1}-1\right)/d \right]$ [dashed lines in Fig.~\ref{fig:results_prl}(a-c)]. 

As demonstrated by Fig.~\ref{fig:results_prl}(a-c), $\mathcal{R}>1$ always. In particular, the full lines in panel (b) characterized by $\omega_1 = 0$ represent the worst scenario, as the corresponding auxiliary state $\ket{\phi_0} = \ket{11}$ is minimally sensitive to dephasing. Even in this case, the probabilistic protocol yields an advantage, that increases with $\omega_1$.
This is shown in panel (c), where we vary $\omega_1$, with higher ones corresponding to better values of $\mathcal{R}$.

The post-selection probability $P$ in panel (a) suggests that the probabilistic scheme is beneficial when it is applied once to a single, large computation $U$. In the opposite scenario, i.e., multiple protocol applications to several gates within $U$, the quasi-deterministic scheme is more suitable. In Fig.~\ref{fig:results_prl}(d), we consider the $T = e^{i\frac{\pi}{8}Z}$ gate and set the desired post-selection probability threshold to one, i.e., no outcome is ever discarded.
As it is possible to see, even in the completely deterministic case {\prot} is advantageous, particularly for large $\omega_1$. Owing to the optimization required for the quasi-deterministic scheme, it is more suited to enhancing several low-fidelity gates (e.g., entangling ones) within larger computations. 

A relevant question to address is how detrimental is the noise affecting the cSWAPs at steps~\ref{prot:step2} and \ref{prot:step4} of {\prot}. This is investigated in Fig.~\ref{fig:results_prl}(e), where we consider $d=2$ branches and $U = [\text{cNOT}(T\otimes T)]^{N_{\rm L}}$, i.e., $N_{\rm L}$ layers of two T gates followed by a cNOT. Instead of the GB model (as for the previous numerical results) here we employ MB-QC (see Fig.~\ref{fig:compstandard1_prl}). Noise is implemented via depolarizing channels applied onto \textit{each} qubit within the resource state, before the measurements (therefore, it affects both the cSWAPs and $U$).

As demonstrated by Fig.~\ref{fig:results_prl}(e), even with noisy cSWAPs our protocols are advantageous, provided $N_{\rm L}$ is large enough. Furthermore, the post-selection probability $P$ is always more than 50\% of the incoherent fidelity for the values of $p_{\rm ne}$ and $N_{\rm L}$ shown here. Qualitatively, it indicates that $\mathcal{R} \geq 1$ is achieved when the noise affecting $U$ is comparable to or larger than the one affecting the cSWAPs. This is particularly appealing in view of recent theoretical and experimental proposals \cite{Levine2019,Gu2021, Kim2022} for high-fidelity multi-qubit gates. 

Finally, in  Fig.~\ref{fig:results_prl}(f) we study the performance of the SQEM protocol using the IBM Quantum Platform. We consider the case $U=\id$  subjected to a dephasing (orange) or depolarizing (brown) channel with error rates $1-p_{\rm ne}$. The dots are reconstructed from the experimental density matrices, which are obtained from tomography after readout mitigation. The dephasing and depolarizing channels are effectively implemented by running circuits with different combinations of Kraus operators, and adding up the measurement outcomes of these circuits weighted by the occurrence probability of their associated Kraus operators. The cSWAP gates, also realized on the real hardware, consist of several cNOT and single-qubit gates. To reconstruct the expected behavior of the IBM hardware, we model each cNOT as a perfect operation followed by a depolarizing channel, with the error probability estimated from the state tomography. We then feed these parameters into our simulator to find the solid lines, which are in good agreement with the dots.

As demonstrated by Fig.~\ref{fig:results_prl}(f), despite the extremely faulty cSWAP operations there is a wide window for which SQEM is advantageous compared to the incoherent case. Importantly, this also holds for the depolarizing case, for which $\omega_1 < 1$ [see Eq.~\eqref{eq:omegas_prl}]. SQEM performance in the small error regime could be dramatically enhanced by better implementations of the cSWAP gates, see, e.g., \cite{Monz09,Reed2012,Levine2019,Gu2021}.

\textit{Conclusions.---} 
We have introduced protocols for quantum noise mitigation that rely on a coherent implementation of a desired computation. Analytical derivations shed light on the working principles of our schemes, and numerical and experimental (IBM Quantum) simulations showcase a significant advantage in computational fidelity under a broad range of settings. 

In \cite{pra_us} we provide additional studies on the feasibility of the protocol. We consider practical auxiliary states and avoid hidden resources. Moreover, we introduce both a nested strategy for further enhancing the fidelity, and a so-called coherent quantum memory that uses our protocols to improve its coherence time. Finally, we propose a physical realization called interferometric-based, which is based on similar working principles albeit not requiring auxiliary states. In this case, correlations between the input and the environment (vacuum) are generated, and the resulting fidelity depends on ``vacuum phases'' between the noisy channels affecting the input in different branches. 

The tools and ideas introduced in this work are not only limited to enhancing quantum computations. They hold the potential to impact multiple fields that are related to quantum information processing, such as quantum communication, metrology or sensing.

\textit{Acknowledgments.---}
This work was supported by the Austrian Science Fund (FWF) through projects No. P36009-N and No. P36010-N. Finanziert von der Europäischen Union - NextGenerationEU. Furthermore, we acknowledge support from the Natural Sciences and Engineering Research
Council of Canada (NSERC), the Canada First Research Excellence Fund (CFREF, Transformative Quantum Technologies), New Frontiers in Research Fund (NFRF), Ontario Early Researcher Award, and the Canadian Institute for Advanced Research (CIFAR). LD acknowledges the EPSRC Quantum Career Development grant EP/W028301/1. We acknowledge the use of IBM Quantum services for this work.

\textit{Additional comments.---}
After completing this work, we became aware of a similar approach independently put forward in \cite{Gideon22,Hann_19}.

\bibliographystyle{apsrev4-2}
\bibliography{bib}

\end{document}